# A new reasonable scenario to search for ER-alpha energy-time-position correlated sequences in a real time mode.

Yu.S.Tsyganov

tyura@sungns.jinr.ru

*FLNR, JINR 141980 Dubna, Russia*

**Abstract**

*A new real-time PC based algorithm and a compact C++ code to operate in a real-time mode with a 48x128 strip double side position sensitive large area silicon radiation detector Micron Semiconductors (UK) are developed and tested. Namely with this new approach it has become possible to provide the quick extraction of EVR-alpha correlated sequences in heavy ion induced complete fusion nuclear reactions. Specific attention is paid to the application of new CAMAC 4 M modules for charge particle position measurement during long- term experiments aimed to the synthesis of new superheavy nuclei. Some attention is paid to the different (combined) algorithm scenario to search for ER-alpha and alpha-alpha chains.*

1. **Introduction**

The Dubna Gas Filled Recoil Separator (**DGFRS**) is the most effective facilities in use for the synthesis of super heavy elements (**SHE**) [1]. Using this facility it has been possible to obtain 49 new super heavy nuclides for about fifteen last years. The PC based detection system allows storing event by event data from the complete fusion nuclear reactions aimed to the study of rare decays of **SHE** [2]. The parameter monitoring and the protection system of the separator [3] is applied to provide for operation safety in long- term experiments with high intensity heavy ion beams and highly active actinide targets, like U, Pu, Am, Cm, Bk and Cf, as well as to provide for the monitoring of the experimental parameters associated with the DGFRS, its detection system and the U-400 FLNR cyclotron.

With the development of an "active correlation" method a new epoch starts in the field of detecting the ultra rare decays of superheavy nuclei (**SHN**) [1,2]. It was namely the Dubna Gas Filled Recoil Separator (DGFRS), the facility which was applied for the discovery of new SHN's in $^{48}$Ca induced nuclear reactions [4]. Along with a new DGFRS detection system which was being put into operation in 2010 a new **REDSTORM** C++ Builder code (**RE**al-time **De**tection and **STOR**age of **Mu**lti-chain events) was designed and successfully applied in the $^{249}$Bk+$^{48}$Ca→117+2-4n complete fusion nuclear reaction [5]. This code is written for two different scenarios. One of them is to operate with the 32-strip position sensitive **PIPS** detector (CANBERRA NV, Belgium), whereas the second one is developed for the 48x128 strip **DSSSD** detector (Micron Semiconductors, UK). The both scenarios of the **REDSTORM** code contain a code fragment which allows searching for a pointer to the potential ER-α correlation sequence in a real-time mode. It means that nearly just after the detection of ER-α the energy-time-position correlated chain code provides the stop of the target irradiation process for a short time in order to detect the forthcoming decays in a background free mode.

2. **New DSSSD based spectrometer design: basic idea**

    As to the specificity of applying DSSSD detector and the development of both real-time algorithm and electronics modules one should keep in mind the following:

    - The detector's structure corresponds to the matrix of the given dimension which, to a first-time approximation, can be used as the matrix of recoil nuclei; its elements are filled in by the value of the current time taken from CAMAC hardware upon receiving corresponding events;
    - Due to the presence of P+ isolating layer between two neighbor strips on the ohmic side of the detector (48 front strips) the edge effects are negligible;
    - On the contrary, for 128 back strips (p-n junction) the effect of charge sharing between the neighboring strips can be up to some 17% in the geometry close to 2π. Certainly, this effect should be taken into account when developing and applying both the algorithm and the electronics modules.

    If briefly, one of the main ideas in the present spectrometer design is changing back side strip ADC modules by the "address detection" modules with no signal amplitude converting. Note, that allow to exclude more than eight 16 in ADC's from the detection system and provides no duplication signals readouts. It is the clearer architect of the event for programmer. The bloc-diagram of the whole process is shown in the Fig.1.

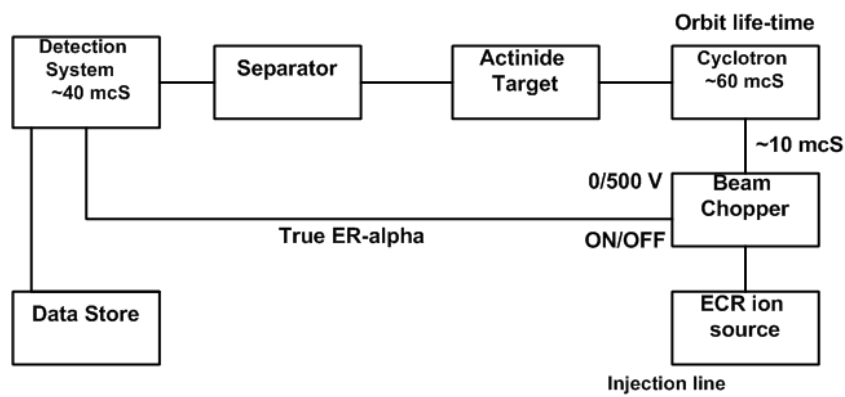

Fig.1 Block-diagram of the real-time process

A brief new idea of the spectrometer design is of no use to ADS's back strip signal processing. Except that, special 4M CAMAC based electronic module is designed [6] to determine only strip number. Seven bit information is written into 16 bit state register. Additional bits are related with other modules in the CAMAC crate. Note, that CAMAC standard is used in order to operate together with the ultra fast digital electronics which, from the other hand, has no possibility to search for ER-α sequences in a real-time mode due to the absence of fast floating point operations during a conventional event by event data acquisition process. The block diagram of such data acquisition is shown in the Fig.2. Note, that along with this system it has become possible to provide strict parallelism in two main tasks processing, namely: a) ultra fast data collecting ( ≈500 nS dead time per event – digital (ORNL) electronics, PIXIE-16 ) and b) searching for ER-α sequences ( ≈10 μS dead time ; CAMAC).

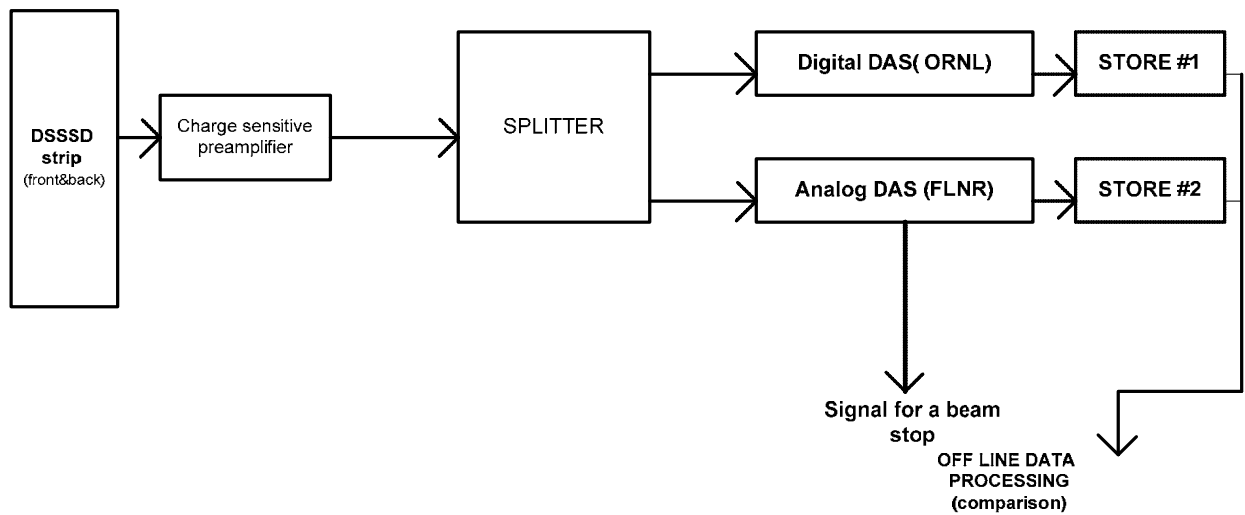

Fig.2 Block diagram of the spectrometer of the DGFRS

3. **Equation's system to provide a choice between two different scenarios for a background suppression**

There are two reasonable scenarios to provide background suppression via cyclotron beam stop when one detects a signal (or/and combination of signals) indicating that a true multi chain event will follow for short time with some significant probability. One of them is to provide a real-time search for ER-α correlation in a real-time mode, whereas the other one is more trivial- to provide a beam stop just after ER signal detection for a shorter time and to try to detect α-like signal in a beam-off time interval. Comparison for these two approaches is considered in details in the Ref. [7 ].

In the present paper the result from [7] is written in the form of three equations system which establishes the border between the two above mentioned scenarios, namely:

$$\eta(t_1, t_2, \tau_{ER-\alpha}) \leq \mu \ll 1,$$

$$t_1 \leq \nu_\alpha \tau_{ER-\alpha} t_2,$$

$$P(\max\{t_1, \tau_{ER-\alpha}\}, \tau_0) \geq 1 - \varepsilon,$$

$$\text{Lg } N_b \leq - N_{min}.$$

In these formulae: μ- is the acceptable by the experimentalist's level of the whole efficiency losses, ε << 1 – the small value parameter, P- the probability to detect one decay of nuclide under investigation during $T_{exp}$ time(duration of the experiment) and $\tau_0$ is the theoretically estimated in advance of the lifetime for the nuclide under investigation. $N_b$ is the expectation parameter value for the given multi chain event to be explained by the set of random factors and $N_{min}$ is the accepted one by the experimentalist's level of statistical significance.

In the same paper a rough estimate was done for the realistic rate values of DSSSD detector recoils and alpha particles. It has been shown that only for correlation time less than approximately ten milliseconds a trivial algorithm for a beam stopping process may take place.

The value of the target irradiation time loss was estimated at about 6% for 48x128 the DGFRS strip detector application.

Note, that one can use the combined algorithm [7], that is both ones are actual under definite circumstances. The block-diagram of this process is shown in the Fig.3 schematically.

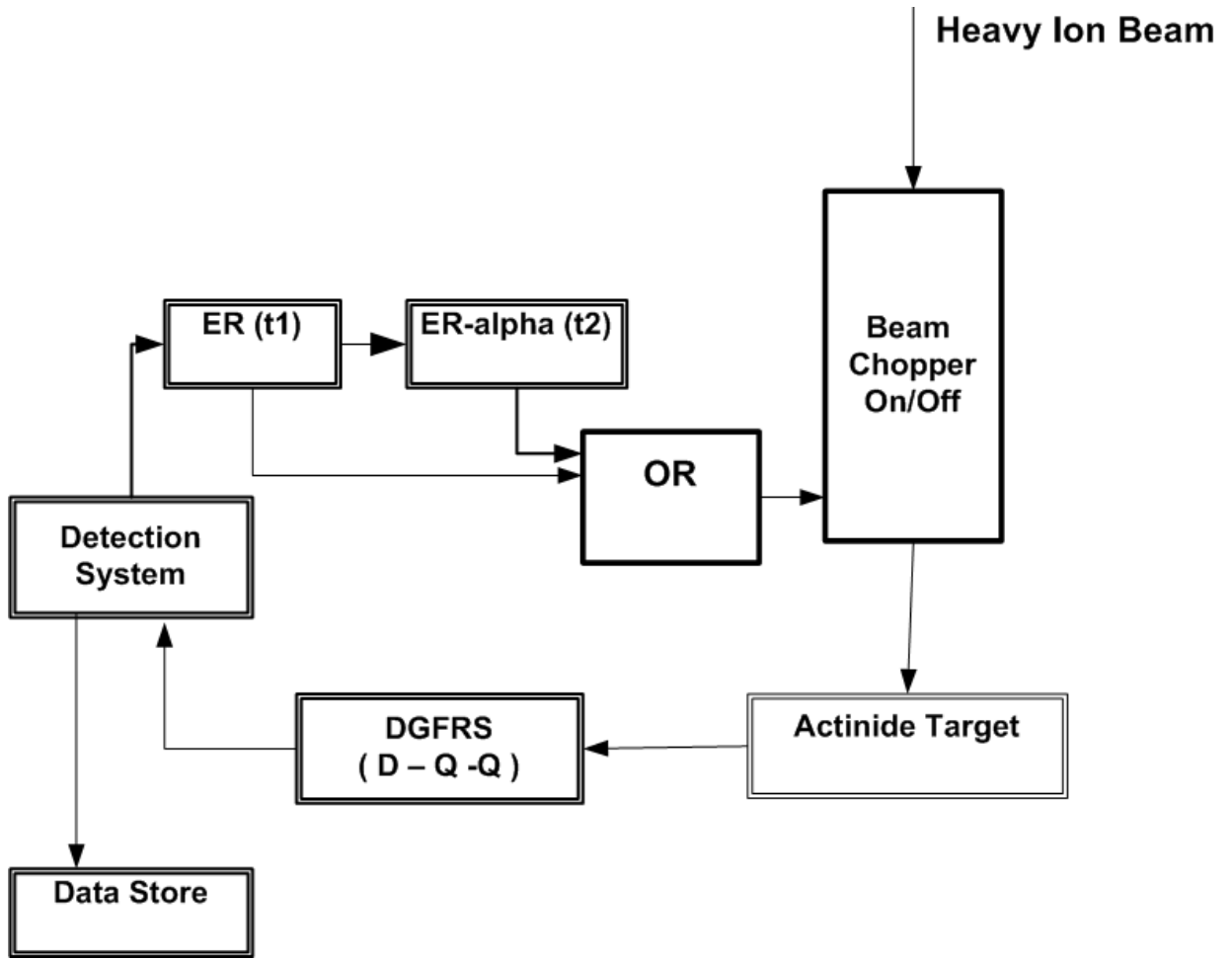

Fig.3 Block-diagram of the combined process [7].

4. **The case of alpha-alpha correlations detection if efficiency of ER detection is not close to 100%.**

Let us consider a case of a few subsequent alpha decay chains when efficiency of ER detection is not close to 100% like it is considered above. The corresponding decay picture is shown in the Fig.4. If one considers 2D picture, Fig.5 except for Fig.4, connects all nodes ((n·(n-1)/2 links in total) by oriented lines and places $\alpha^k_{ij}$ matrixes onto the graph vertex. Here k is the number of the detected signal which can be attributed to alpha-decay of SHE. It is possible to compose for each alpha particle signal candidate the relationship like $\Delta t^{k,k+n}_{i,j}=\min\{\alpha^k_{i,j}-\alpha^{k+n}_{i,j+m}\}|_{m=0,1,-1}$. Hence, if at the given time moment this parameter is less or equal than setting parameter $t_{kn}$, then the system can generate a beam stop for a short time.

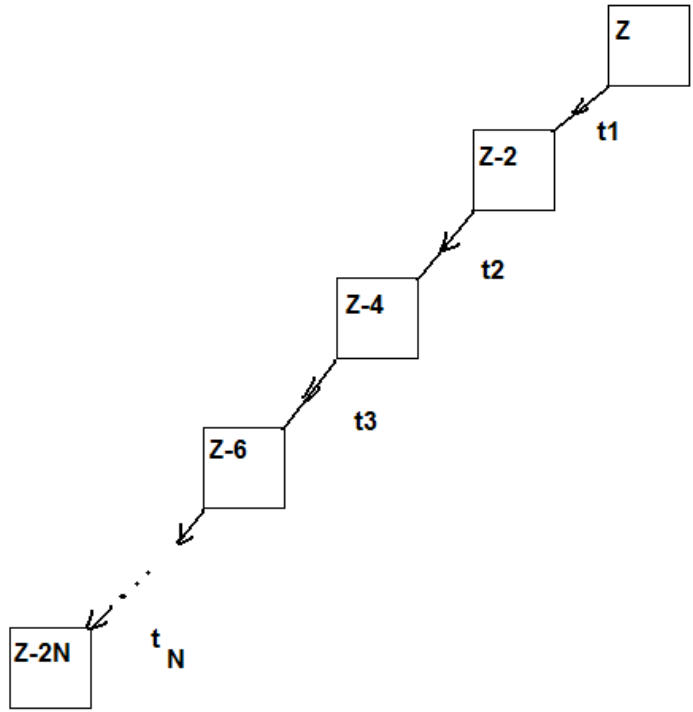

**Fig.4** Alpha decay chains 1…N for Z to Z-2·N.

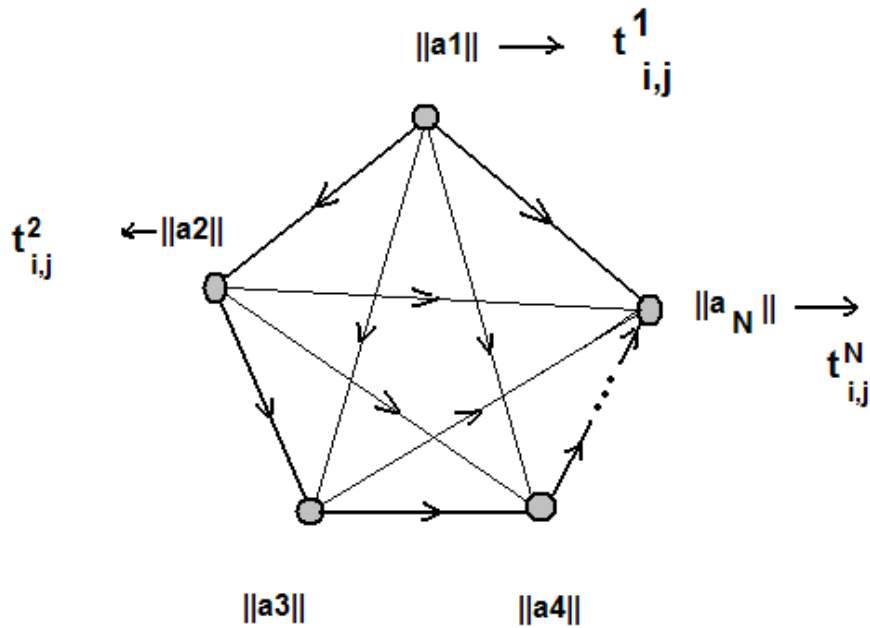

**Fig.5** Schematics of the algorithm for $\frac{N\cdot(N-1)}{2}$ α-α correlated chains.

If, according to the requirements of an experiment beam-stop after missing **$n$** alpha particle is available and $\psi$ is efficiency to detect alpha particle by a focal plane detector, then one can consider the value of $\boldsymbol{P_n = \psi \cdot \sum_{i=0}^{n}(1-\psi)^i}$ as a probability to generate the mentioned beam-stop signal. In the real experiments the parameter of $\psi$ is close to 0.5, although if one takes into account detection not only by focal plane detector, but also by side detector too, then it may be as about 0.7-0.85 depending on the energy threshold of the detection system. And of course, it is easy to include the ER signal into the above mentioned process consideration with the parameter of the detection efficiency $\psi_{ER} \approx 1$.

## 5. Builder C++ code TVPS.exe for electronic modules testing and data acquisition

C++ TVPS.exe code is designed for two general purposes. One of them is testing CAMAC electronics modules like ADC PA [6], "address detection" CAMAC 4M module and state register one(1M). In part, these tests are described in [8].

The second branch of the code application is related with the data taking in the long term experiments aimed to rare alpha decays detection (or/and spontaneous fission). To a first approximation it planned to use no any module for elapsed time value. Except that, one may use the internal time with microsecond accuracy using the example procedure presented below.

*// example – the time measurements of delta () function execution in 100 cycles*

*LARGE_INTEGER StartingTime, EndingTime, ElapsedMicroseconds;*

*LARGE_INTEGER Frequency;*

*QueryPerformanceFrequency(&Frequency);*

*QueryPerformanceCounter(&StartingTime);*

*for ( int j=0; j <100 ; j++) delta();*

*QueryPerformanceCounter(&EndingTime);*

*ElapsedMicroseconds.QuadPart=EndingTime.QuadPart-StartingTime.QuadPart;*

*ElapsedMicroseconds.QuadPart \*=1000000;*

*ElapsedMicroseconds.QuadPart /=Frequency.QuadPart;*

*Form1->Caption=ElapsedMicroseconds.QuadPart;*

*//Form1->Caption=double(EndingTime.QuadPart);*

The flowchart of the process and the schematics of ER matrix element formation are shown in the Fig.6a and Fig.6b respectively.

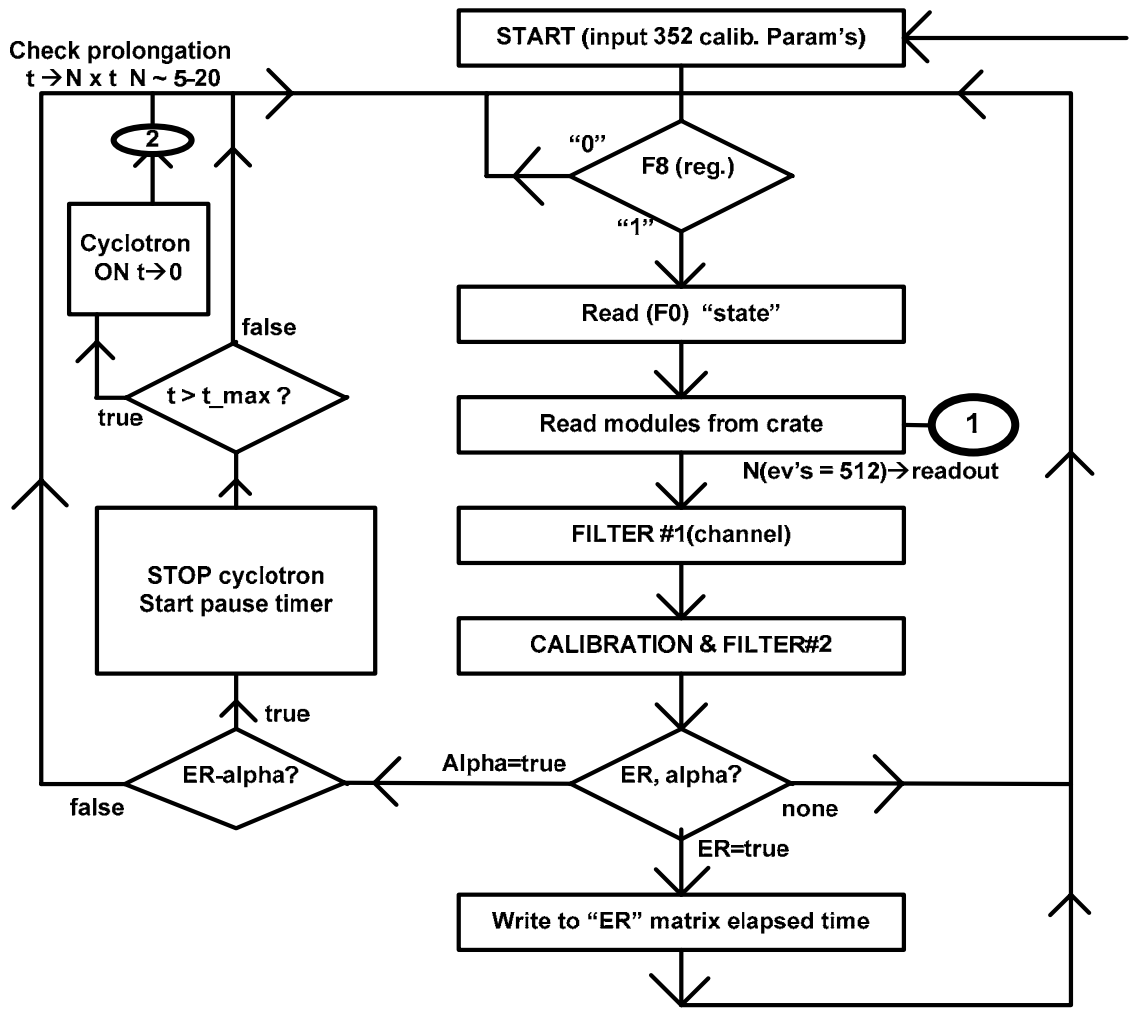

Fig.6a The flowchart of the process data taking and ER-alpha searching ( 1- check beam off pause prolongation; 2- file writing ).

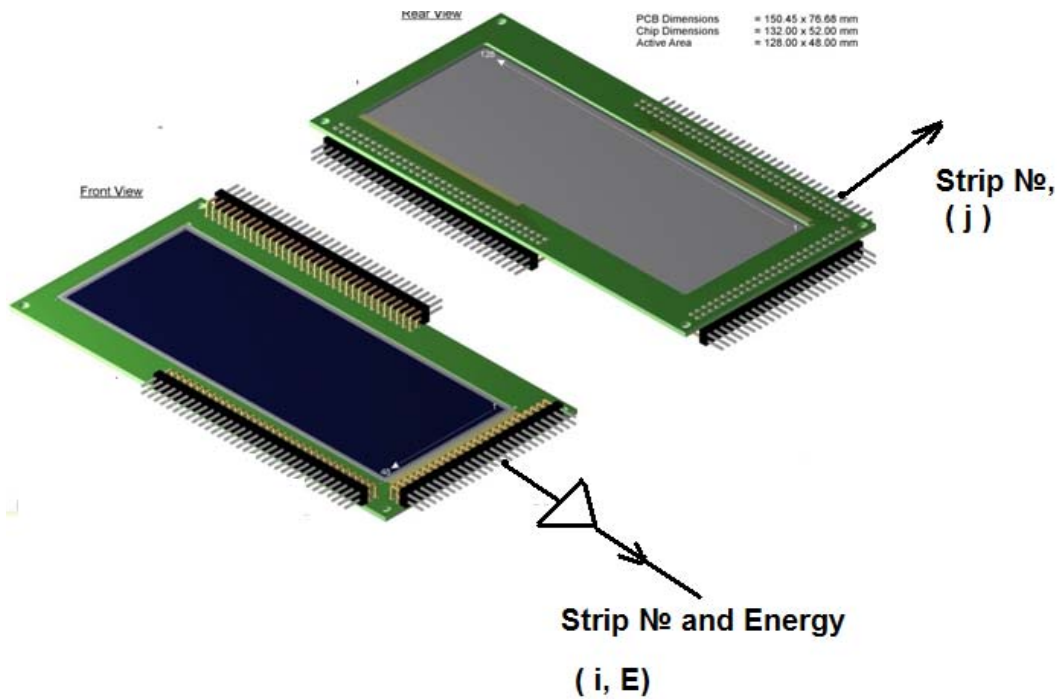

Fig. 6b Schematics of ER matrix element (ER elapsed time $t_{i,j;\ ER=true}$) formation. View (front and rear) of DSSSD detector of the DGFRS

As it is presented, the FILTER#1 routine provides discrimination according to the channel number in the Fig.6a flowchart, whereas the FITER#2 routine provides event discrimination according to a minimum energy level. For filling a ER matrix element with an elapsed time the minimum and maximum values both for energy and time-of-flight values are taken into account. An example of typical value of "STATE" parameter (binary, 16 bit) is as following: 1001101010111100. Here:

- First seven bits – the code of back side strip;
- next three bits – the code of the front strip ADS number or the code of side ADC;
- next four bits (1111) denote that all three amplitudes (TOF,$\Delta E_1$, $\Delta E_2$) and the mark of fly are non zero values;
- 15$^{th}$ bit (0) is the majority coincidence mark;
- 16$^{th}$ bit – reserved, not in use.

For the sake of calibration procedure one usually applies the complete fusion nuclear reaction

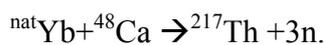
$^{nat}$Yb+$^{48}$Ca → $^{217}$Th +3n.

The transmitting both "beam off" and "beam on" TTL signals is performed through the DGFRS control system described in [ 3 ]. As to the choice of minimum/maximum parameter of ER registered energy value the systematic "calculated incoming energy- registered energy" from [9 ] is used.

The simplest approximation for the ER mean energy can be presented by the formula:

$$E(reg) \approx -1.7 + 0.74 \cdot E(in), \ [\text{MeV}].$$

Here, E ( reg ) – the registered value and E(in) – the incoming one, respectively.

Of course, for the same purpose, the PC based numerical simulation reported in [10 ] is useful too.

The project TVPS.bpr contains two main Builder forms. One of them is presented in the Fig.5 serves for electronics modules tests, whereas the second one corresponds to the main event by event acquisition mode for long-term experiments. The menu item for acquisition in a event by event mode (and to open Form2, respectively) is shown by an arrow in the Fig.7. The two different colors (red and grey) of the Form2 middle area indicate whether "beam stop" mode is actual or not at a given time. In the case of "beam stops" the mode of application is actual edge effects for back side strips are taken into account in the manner reported in [1].

The right side columns show elapsed times [µS] as well as (right) the time difference between the two incoming events. Table 1 shows automatically founded peak center positions for two peaks and event rate value (element 4, 1). In the left-down window it is shown the whole acquisition time value. The position of the ADC under test is equal to 9 (right-bottom).

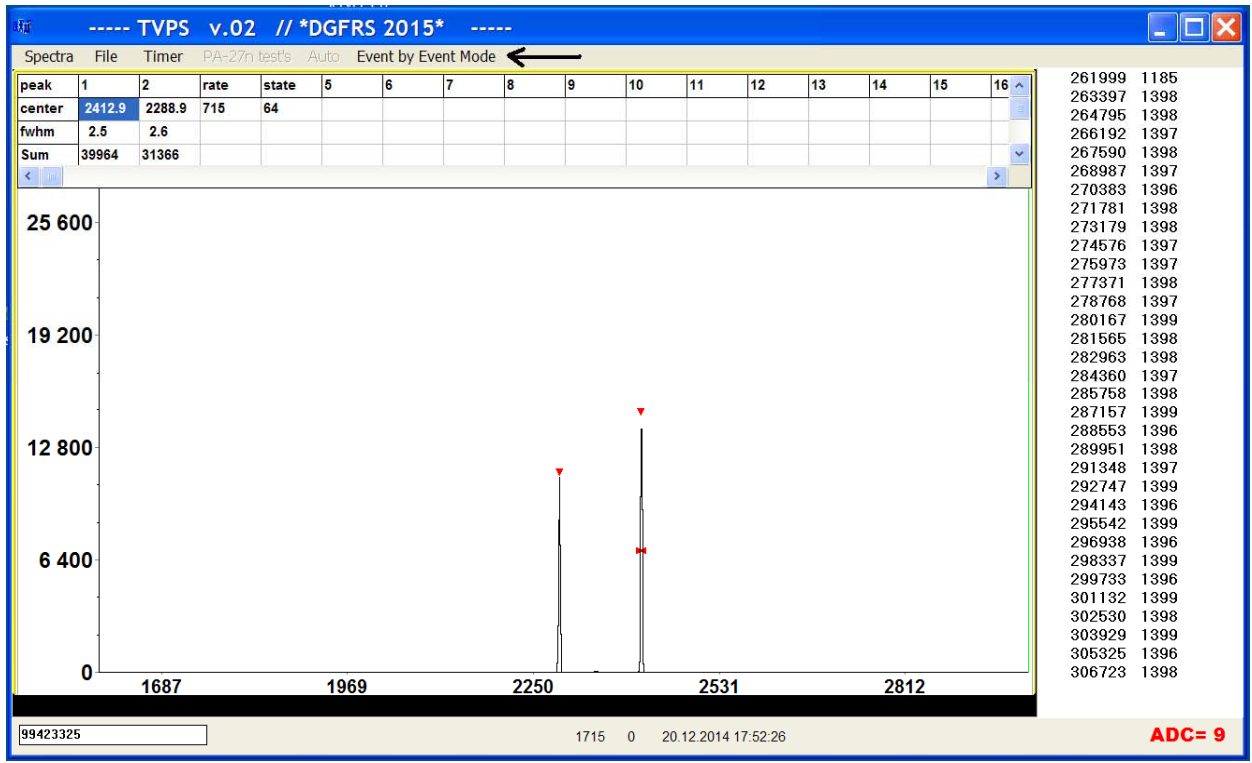

Fig.7 The main window of TVPS code.

6. **Summary**

C++ Builder PC based (Windows) application TPVS is designed for heavy ion induced complete fusion nuclear reaction experiments aimed to the synthesis of SHE. Simple "non-beam" tests are performed. This application is designed in order to operate together with a new electronics and DSSSD detector.

In a nearest future it is planned to provide extensive tests at U 400 FLNR cyclotron $^{48}$Ca beam. Additionally to the system described in the present paper, the new version of the DGFRS safety and monitoring system development is in progress now too and will put into operation in 2015. Along with that system putting into operation the application described in the present paper C++ TPVS code will be definitely more effective. The paper concludes a set of papers [8,9,11,12,13] connected with the automation process of the DGFRS experiments. This paper is supported partly by the RFBR grant №000.

**Supplement 1. A few words about calibration process**

The heavy ion induced complete fusion reaction $^{nat}Yb+^{48}Ca \rightarrow$ Th+xn is usually used to calibrate both PIPS (resistive strip) and DSSSD detectors at the DGFRS. Typically it takes a few days to provide thoroughly calibration procedure and to obtain a hundreds of parameters as a result. Recently, different techniques are developed to simplify the whole calibration process. The method, reported in [12] is one of them. The general idea is of using 9.26 MeV line ($^{217}$Th) as a first "one peak" approximation. In the next step two additional peaks are used in a form of three peaks least square method application. A more universal and extended method is reported in [13,14]. This approach uses quasi-curvature parameter to find peaks position for the calibration process.

**Supplement 2. High beam intensities applications development**

Along with commissioning in a nearest future very high intensity DC-10 cyclotron [15] beam intensity value (e.g. $^{48}$Ca projectiles) will reach approximately up to 10 pµA at the actinide target position. Of course, some precaution should be made in order to provide non-destruction operation with highly radioactive target. From the other hand, the application of "active correlations" method will be required to provide the high level of statistical significance for the detected multi chain event and some problems will definitely arise with that application due to the higher rate of events at the focal plane of the DGFRS in comparison with U-400 cyclotron application. The author does not exclude that some additional requirements to the ER-α parameters are going to place. Definitely, one of them is the estimated level of probability of ER-α chain to be a random. In that case, C++ code will contain restriction for that probability in order to minimize the losses of the whole experimental efficiency. Of course, this calculation of the ER-α random coincidence probability value should be performed strongly in a real-time mode. Extra parameters for ER identification, like ΔE signals (in addition to TOF signal) both for START and STOP counters are welcome. A larger strip number for DSSSD detector is not excluded to provide a better positional resolution for ER-α correlation link. In that connection a general form of required condition for detection a correlation sequence will be as:

$$E_{ER} \in \left(E_{ER}^{min}, E_{ER}^{max}\right) \;\&\&\; TOF \in \left(TOF_{ER}^{min}, TOF_{ER}^{max}\right) \;\&\&\; \Delta E_{ER}^{start} \geq \Delta E_{min} \;\&\&\; \Delta E_{ER}^{stop} \geq \Delta E_{min}$$

$$E_{\alpha} \in \left(E_{\alpha}^{min}, E_{\alpha}^{max}\right) \;\&\&\; TOF = 0 \;\&\&\; \Delta E_{\alpha}^{start} < \Delta E_{\alpha}^{min} \;\&\&\; \Delta E_{\alpha}^{stop} < \Delta E_{\alpha}^{min}$$

$$P_{CORR}^{i,j\pm 1} \leq \varepsilon \ll 1$$

$$\Delta t_{ER-\alpha} \leq \tau_0 \;\&\&\; POSITION_{ER}^{XY} \approx POSITION_{\alpha}^{XY}$$

Here, indexes ER, α are corresponded to recoil and alpha particle signals, respectively and start/stop – to START and STOP proportional chamber signals, respectively. Parameter $P_{CORR}^{i,j\pm1}$ denotes the correlation probability value ( i, j- strip number for front and back side signals,respectively) , mentioned before, and ε > 0 is an infinitesimal value, POSITION is a XY detected with DSSSD detector position, $\tau_0$ is a pre-setting time parameter. Sign $\pm 1$ is used due to edge effect of charge sharing between two neighbor back strips. It is important that all calculations are performed with taking into account a local intensity value (average one for a few minutes).